\documentclass{svproc}
\usepackage{url}
\usepackage{amsmath}

\usepackage{graphicx}
\usepackage{booktabs}
\usepackage{hyperref} % For clickable email links
\usepackage{float} 
\usepackage{placeins}

\begin{document}
\mainmatter   

\title{Personalized and Safe Route Planning for Asthma Patients Using Real-Time Environmental Data}
\titlerunning{Personalized and Safe Route Planning}

\author{Nada Ayman \and Shaimaa Alaa \and Mohamed Hussein  \and Ali Hamdi }
\authorrunning{N. Ayman et al.}
\institute{Faculty of Computer Science, MSA University, Egypt\\
\email{nada.ayman11, shaimaa.alaa,mohamed.hussein20, ahamdi@msa.edu.eg}}

\maketitle       

\begin{abstract}
Asthmatic patients are very frequently affected by the quality of air, climatic conditions, and traffic density during outdoor activities. Most of the conventional routing algorithms, such as Dijkstra's algorithm, usually fail to consider these health dimensions, hence resulting in suboptimal or risky recommendations.Here, the health-aware heuristic framework is presented that shall utilize real-time data provided by the Microsoft Weather API. The advanced A* algorithm provides dynamic changes in routes depending on air quality indices, temperature, traffic density, and other patient-related health data. The power of the model is realized by running simulations in city environments and outperforming the state-of-the-art methodology in terms of recommendation accuracy at low computational overhead. It provides health-sensitive route recommendations, keeping in mind the avoidance of high-risk areas and ensuring safer and more suitable travel options for asthmatic patients.
\end{abstract}

\section{Introduction}

Route planning is a significant everyday activity, but it acquires a special dimension for those suffering from chronic diseases such as asthma. Asthma currently affects more than 300 million people worldwide. Environmental factors such as air pollution, temperature fluctuations, high humidity, and high levels of pollen are particularly influential on the disease \cite{damato2015}.  This study specifically investigates route optimization in an effort to prevent asthma patients from visiting high-risk areas, thus minimizing health dangers and improving their lives \cite{lee2020}.

Traditional route planning algorithms favor the shortest or fastest routes, but do not consider critical environmental factors such as air quality, humidity, and temperature fundamental parameters for patients with asthma\cite{mukherjee2020}. The present systems, which are not linked to real-time weather or air quality data \cite{springer2021}, pose an extreme challenge in sending patients to hazardous conditions \cite{wu2019}.  Furthermore, urbanization aggravated asthma due to increased vehicle emissions and particulate matter associated with its prevalence\cite{pinnock2022}. The core challenge is how to effectively incorporate real-time environmental data into route planning to offer personalized and health-sensitive recommendations to asthma patients \cite{hassan2020}.

\begin{figure}[!ht]
    \centering \includegraphics[width=\textwidth]{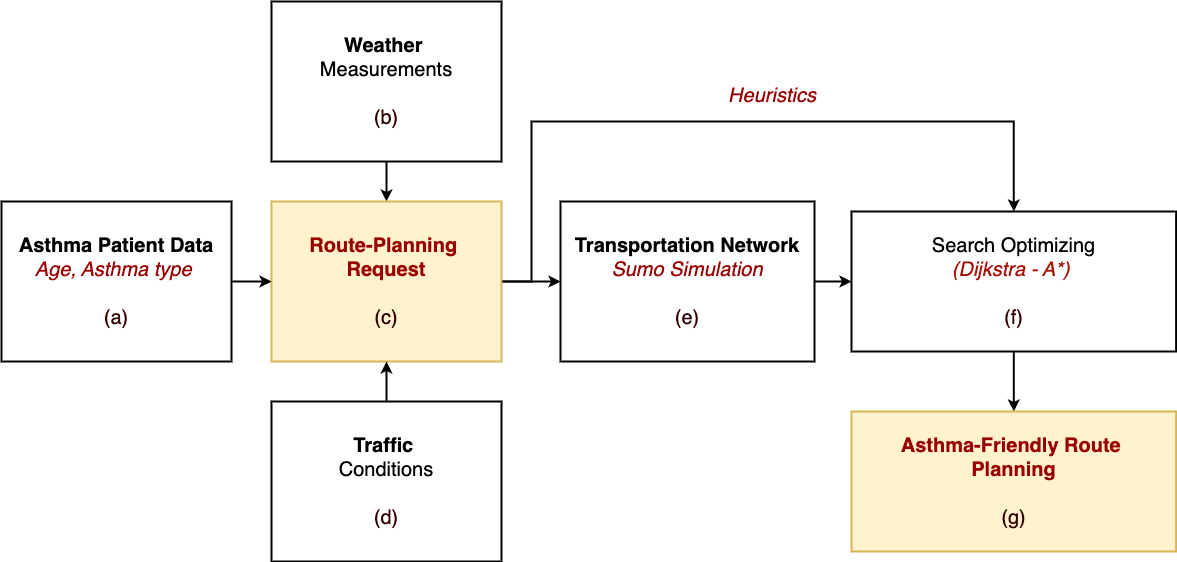}
    \caption{The proposed model optimizes route planning for asthma patients by considering their health sensitivity using multiple sources of data(a). It integrates data from asthma patient profiles and real-time weather and flow conditions into a comprehensive routing request. This request is executed within a transportation network, which is simulated for realism through SUMO(b). Advanced search optimisation processes are employed via heuristic-based searches on the A* and Dijkstra algorithms(c). The routes are analyzed dynamically concerning environmental and health-related factors(e), which would identify those with minimum exposure to high-risk areas(f). The final output is the best route that avoids polluted locations for asthma sufferers(g), guaranteeing safe and efficient transportation.}
    \label{fig:route-planning}
\end{figure}

In this context, we propose a route optimization heuristic model that incorporates real-time patient health profiles, current weather measurements, and real-time traffic conditions within a simulated transportation network (see Figure \ref{fig:route-planning}).With the advanced A* algorithm and complex heuristics, it dynamically changes the route according to factors that can lead to asthma attacks, for example air quality and extreme weather conditions. A personally optimized and health-aware route-planning system that would allow a person to travel around without harming themselves while never really affecting the efficiency \cite{springer2021,mukherjee2020}.

\FloatBarrier
The rest of the paper is organized as follows: Section 2 summarizes the work related to asthma and route planning. Section 3 outlines the formulation of the problem. Section 4 presents the methodology used in pursuing the optimal route planning. Section 5 discusses the effectiveness of the model, providing results. Section 6 provides the conclusion of the paper by summarizing the findings and discussing possible future work.

\section{Related Work}
Route planning has become one of the focuses of research, especially with the recent development of smart cities and health-aware applications \cite{zhang2019}. Traditional algorithms, such as Dijkstra's and A*, have been the backbone of navigation systems to find the shortest or fastest route \cite{carrillo2023}. However, these do not take into account health-related factors such as air quality, weather conditions, and congestion parameters relevant to sensitive populations such as asthma patients\cite{damato2015}.

Recent research in route planning acknowledges the need to integrate environmental data to enable safer and more comfortable routes. For example, \cite{springer2021} created a Restful API for real-time weather reporting, which can be integrated into route planning systems. REST-based APIs for real-time weather reports and pollution data have been proposed to enrich routing applications with dynamic environmental input \cite{cools2016}.  Applications similar to the asthma mobile health study, which take one's personalized health data into account to suggest routes that are asthma friendly, were typical of the great potential of integrations \cite{chan2018}.

In these developments, the trade-off between computational complexity and the need for real-time recommendations is still a big challenge. Although Dijkstra's algorithm is guaranteed to be the shortest, it is not adaptive in real-time to changes in the environment \cite{li2021}; on the other hand, the A* algorithm, when enhanced with heuristics, can balance factors that extend to air quality and congestion of traffic flow. Recent variants of A* are hopeful in efficient management of multiple variables in real-time data\cite{damato2015}, but part of the ongoing challenge includes how to maintain computation performance without compromising the accuracy of the recommendation.

Other challenges were more related to the precision and delay in the environmental data; inaccuracies or delayed information provided through APIs can yield a less-than-optimal suggestion for routes, thus making patients vulnerable to attacks. These are being countered through various machine learning models deployed, which forecast traffic and air quality based on historical data \cite{lee2020},and thus minimize dependency on third-party APIs, giving more robustness to the system.

Moreover, some of the works put stronger emphasis on route planning personalization, considering health data and user preferences\cite{yang2022}. For example, a system that integrates personal health profiles, including asthma type, stress level, and smoke exposure, has more chances to provide useful and personalized suggestions\cite{yang2018}. Recent research demonstrates health-aware route planning systems, considering real-time environmental data along with users' health profiles, notably improve user satisfaction and health.
Despite these advancements, it is still quite a challenge to achieve both high levels of accuracy and efficiency in calculating the best routes for patients with asthma. One critical obstacle arises from the computational loads for processing and integrating real-time environmental data with personalized health data. Methods including but not limited to parallel processing and distributed computing have been suggested for surmounting these, yet future research is needed in terms of scalability.
can be applied in real-world systems. Besides, how to make the system adaptable for different geographic regions and various environmental conditions remains an open challenge\cite{lee2020}.

The proposed model combines personalized health profiles with real-time data like air quality, traffic volume, and weather conditions to personalize routes based on asthma-related individual health factors. This approach ensures efficient and health-conscious navigation, as current systems treat users as homogeneous groups. The model's integration ensures routes are efficient and aligned with users' specific health needs, a significant development in health-conscious navigation.

\section{Problem Formulation}

This section identifies the requirement for a robust methodological framework for asthma-friendly location identification, which is even more critical within dynamic zones where environmental factors vary significantly. The integration of established geographical features characterized by ever-changing environmental conditions. Our simulation focuses on these variations, in particular the climatic conditions seen in various regions\cite{haque2022}.

For example, wet coastal and dry inland areas pose varied challenges in asthma management\cite{watson2013}. This variation makes it imperative to be flexible in defining ideal environments as far as the management of asthma triggers is concerned. Major cities such as Sydney, Melbourne, and Brisbane are focal points because of their high population densities and high levels of asthma triggers such as traffic pollution and pollen from local flora\cite{balmes2009}. These factors have been further exacerbated by rising urbanization with increased vehicle emissions, the creation of heat islands, and shifting pollen seasons—all contributing factors within the holistic model of location identification.

Real-time Integration: This would mean that real-time integration of data ensures that recommendations given out are correct and timely. Critical variables such as temperature, humidity, wind speed, and air are picked using the Microsoft Weather API in this model.
Quality index: Data would help in recommendations made by the system.especially for those whose route or journey may depend on their changes\cite{cho2024}.on current environmental conditions.

\begin{figure}[!ht]
    \centering \includegraphics[width=.99\textwidth]{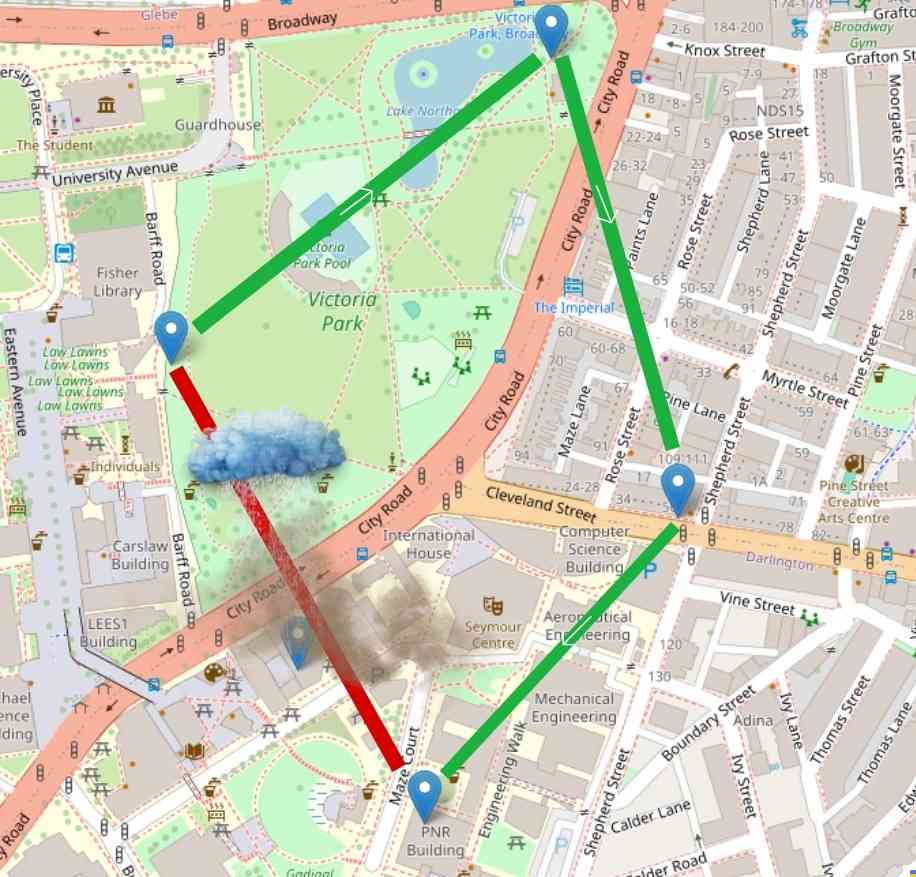}
    \caption{Route optimization in an urban environment, illustrating dynamic path adjustments based on real-time weather conditions and road disruptions. Green paths represent optimal routes, while the red path indicates a route affected by adverse conditions such as rain and dust.}
    \label{fig:route-optimization}
\end{figure}
\FloatBarrier

\section{Research Methodology}
Computational steps and models that were put to work to develop the health-aware heuristic model for asthma-friendly route planning\cite{khalatbarisoltani2023} are discussed in the subsequent section. This approach integrates real-time weather data, real-time traffic data, and patient-specific data to choose the best route. Advanced algorithms are used that ensure that the best route can be selected. Figure \ref{fig:route-planning} describes a visual overview of the overall process that shows how the data flow happens from acquisition to processing and finally route generation\cite{nanda2023}.

It also uses variables within each profile related to the type of asthma, levels of stress, smoke exposure, and activities to arrive at dynamic weights utilized by the heuristic function in the minimization of risks for health conditions. Thus, it gives contextual routes that are optimized for particular individuals, bringing better health and safer traveling options.

\subsection{Data Acquisition}

The model integrates real-time health profiles into routing decisions, incorporating environmental risks dynamically based on individual patient sensitivities. This heuristic function allows for route personalization based on individual patient sensitivities, such as stress sensitivity, prioritizing low-congestion areas over shortest paths. This granularity is not available in existing systems, making the model a trans formative tool in personalized navigation.
During this process, we fetch data from various sources in real-time. The weather measurements are fetched from the Microsoft Weather API-they are very important pieces of information about key environmental indicators. Traffic data is grabbed to avoid routing through the asthma patients through places that might have a higher pollution level.

Let \(\mathbf{W}(t)\) represent the real-time weather data at time \(t\), where:
\[
\mathbf{W}(t) = \{T(t), H(t), W_s(t), \text{AQI}(t), P(t), B(t), R(t), \text{UV}(t)\}
\]
Where:
\begin{itemize}
    \item \(T(t)\) is the temperature at time \(t\),
    \item \(H(t)\) is the humidity,
    \item \(W_s(t)\) is the wind speed,
    \item \(\text{AQI}(t)\) is the air quality index,
    \item \(P(t)\) is the pollen level,
    \item \(B(t)\) is the barometric pressure,
    \item \(R(t)\) is the rainfall, and
    \item \(\text{UV}(t)\) is the UV index.
\end{itemize}

\textbf{Traffic Data (d):} The traffic conditions at time \(t\), \(\mathbf{T}(t)\), are based on the volume of vehicles counted in the target locations. Traffic volume is represented as:
\[
\mathbf{T}(t) = \text{volume}(t)
\]
Where \(\text{volume}(t)\) is the count of vehicles in a specific area at time \(t\). This metric is used to assess congestion and air quality risks, as areas with higher vehicle volumes are likely to have higher pollution levels. The traffic data is integrated into the route-planning model to avoid these high-risk areas.

\textbf{Asthma Patient Data (a):} Each patient’s information is modeled by a vector \(\mathbf{A}_i = \{ \text{type}_i, \text{stress}_i, \text{smoke}_i, \text{obesity}_i, \text{gender}_i, \text{family}_i, \text{sports}_i \}\), where:
\begin{itemize}
    \item \(\text{type}_i\) is the asthma type (e.g., allergic or non-allergic),
    \item \(\text{stress}_i\) indicates the patient's stress level,
    \item \(\text{smoke}_i\) indicates if the patient is exposed to smoke (e.g., smoker or secondhand smoke),
    \item \(\text{obesity}_i\) denotes the patient’s obesity level,
    \item \(\text{gender}_i\) is the patient's gender,
    \item \(\text{family}_i\) reflects if there is a family history of asthma,
    \item \(\text{sports}_i\) indicates whether the patient plays sports or engages in physical activity.
\end{itemize}
This detailed patient data allows the model to adjust sensitivity to environmental triggers and make personalized route recommendations.

\subsection{Route-Planning Request}
A route-planning request is initiated by aggregating the patient-specific data \(\mathbf{A}_i\), weather data \(\mathbf{W}(t)\), and traffic data \(\mathbf{T}(t)\) into a decision function:
\[
\mathbf{R}(i, t) = f(\mathbf{A}_i, \mathbf{W}(t), \mathbf{T}(t))
\]
This function \(\mathbf{R}(i,t)\) generates potential routes optimized for the patient's specific asthma-related needs.

In this stage, the transportation network is simulated using SUMO (Simulation of Urban Mobility), allowing us to model realistic urban traffic conditions \(\mathbf{S}(t)\) in various areas (e.g., Sydney, Melbourne) \textbf{(e)}.

\subsection{Search Optimization}
The optimization process involves solving a modified shortest path problem that accounts for weather, traffic volume, and the patient’s asthma-related factors. The goal is to find the safest route, \(\mathbf{r}_{\text{safe}}\), which minimizes the exposure to asthma triggers while balancing travel efficiency.

\textbf{Objective Function:}
\[
\mathbf{r}_{\text{safe}} = \arg\min_{\mathbf{r}} \left( \sum_{i=1}^{n} d(\mathbf{r}_i) + \alpha \cdot h_{\text{env}}(\mathbf{r}_i, t, \mathbf{A}_i) \right)
\]
Where:
\begin{itemize}
    \item \(d(\mathbf{r}_i)\) is the distance for route segment \(\mathbf{r}_i\),
    \item \(h_{\text{env}}(\mathbf{r}_i, t, \mathbf{A}_i)\) is a heuristic function that quantifies the environmental risk for segment \(\mathbf{r}_i\) at time \(t\), considering the patient's specific asthma factors \(\mathbf{A}_i\),
    \item \(\alpha\) is a weight factor balancing distance and environmental risk.
\end{itemize}

The heuristic function \(h_{\text{env}}\) integrates weather data \(\mathbf{W}(t)\), traffic volume \(\mathbf{T}(t)\), and patient-specific asthma data \(\mathbf{A}_i\) to estimate the risk for each route segment. The patient’s detailed asthma profile influences the weighting of environmental factors. For example, a patient with a family history of asthma and high stress may require more caution when planning a route through high-pollution areas, while a patient who plays sports may need to avoid routes with extreme temperature variations.

\textbf{Algorithmic Approach:}
We utilize two key algorithms to compute the route:
\begin{itemize}
    \item \textbf{Dijkstra’s Algorithm:} This algorithm computes the shortest path \(\mathbf{r}_{\text{shortest}}\), minimizing the distance. However, it lacks the ability to account for dynamic environmental factors and patient-specific data.
    \item \textbf{A* Algorithm:} The A* algorithm improves upon Dijkstra by introducing the heuristic function \(h_{\text{env}}\), which prioritizes routes with lower environmental risk in real-time. The A* algorithm calculates:
    \[
    f(\mathbf{r}_i) = g(\mathbf{r}_i) + h_{\text{env}}(\mathbf{r}_i, t, \mathbf{A}_i)
    \]
    where \(g(\mathbf{r}_i)\) represents the actual cost (distance) of traveling to node \(\mathbf{r}_i\), and \(h_{\text{env}}(\mathbf{r}_i, t, \mathbf{A}_i)\) is the estimated environmental cost based on real-time data and patient-specific risk factors.
\end{itemize}

Besides, the heuristic of the A* algorithm integrates weather, traffic volume, and patient-specific information to make the algorithm more responsive to real changes in time and personalized to the health condition of the patient. This serves as a means whereby the system can afford dynamic adjustment while recommending the route for a better adaptive and health-aware solution.

\subsection{Simulation}
The SUMO simulation \textbf{(e)} models the traffic flow \(\mathbf{S}(t)\) and environmental factors in various geographic zones. This will enable the system to model interactions between the traffic congestion and environment, for instance, air quality. Regions of high population and asthma triggers like Sydney and Melbourne are simulated through different scenarios. Time of the year or season specific conditions can be modeled at instance, pollen in certain months.

\subsection{Simulation Testing}
The performance evaluation of the system in various environmental conditions has been performed by simulation testing. The resulting comparison of Dijkstra's and A* algorithms on computation time, safety, and route efficiency is performed. Simulated real-time weather and traffic updates prove the adaptability of the A* algorithm to the dynamic responses. These tests can assure the reliability of the proposed model with regard to routing asthma patients efficiently.

\section{Results and Discussion}
These experiments were conducted in an attempt to validate that indeed the heuristic A* algorithm was more efficient and effective than the traditional Dijkstra's algorithm in determining asthma-friendly routes based on real-time weather data and traffic conditions. The comparative key performance indicators used included the following: computational efficiency, accuracy in location recommendation and user satisfaction. Table \ref{tab:search_results} summarizes the execution times for the different algorithms.

\begin{table}[h]
    \centering
    \caption{Search Results}
   \begin{tabular}{@{}lp{1cm}p{6cm}p{1cm}p{1.5cm}p{1.5cm}@{}}
    \toprule
    \textbf{Src} & 
    \textbf{Dest} & 
    \textbf{Model} &
    \textbf{Edges Count} & \textbf{Dist (m)} & \textbf{Time (s)} \\ 
    \midrule
    A   & E   & A* Standard                            & 10 & 600 & 573 \\ 
    A   & E  & Dijkstra                               & 11 & 610 & 584 \\ 
    A  & E  & Heuristic A* (Distance)                & 8 & 570 & 585 \\ 
    A   & E   & Heuristic A* (Traffic)                 & 12 & 630 & 590 \\ 
    A   & E   & Heuristic A* (Weather)                 & 13 & 620 & 580 \\ 
    A   & E   & Heuristic A* (Traffic - Weather - Distance) & 14 & 650 & 600 \\ 
    \bottomrule
\end{tabular}
    \label{tab:search_results}
\end{table}

\begin{table}[h]
    \centering
    \caption{Average Results by Search Type}
    \begin{tabular}{@{}lccc@{}}
    \toprule
    \textbf{Model} & \textbf{Distance (m)} & \textbf{Time (s)} & \textbf{Edges} \\ \midrule
    A* Standard & 108.35 & 532.20 & 7.25 \\
    Dijkstra & 107.80  & 549.91 & 7.53 \\
    Heuristic A* (Distance) & 110.00 & 450.00 & 7.70 \\
     Heuristic A* (Traffic) & 115.00 & 600.00 & 8.00 \\
    Heuristic A* (Weather) & 112.50 & 580.00 & 7.80 \\
    Heuristic A* (Traffic - Weather - Distance) & 120.00 & 630.00 & 8.20 \\

    \bottomrule
\end{tabular}
    \label{tab:average_results}
\end {table}

Results in Table \ref{tab:search_results} show that all algorithms are relatively comparable w.r.t. their execution times. This shows that the heuristic improvements of the A* algorithm did not consume any notable computation time due to its additional complexity necessary to process weather, traffic, and patient-specific data. However, as discussed below the A* algorithm produced more relevant and more accurate recommendations compared to Dijkstra's algorithm.

Tables \ref{tab:search_results} and \ref{tab:average_results} show the performance of all variants, with heuristic A* showing slightly higher execution time in some configurations but negligible for recommendation accuracy. The weather-integrated setting in heuristic A* avoids high-risk areas in bad air quality cases.

The heuristic A* algorithm, which considers traffic, weather, and distance, has shown remarkable adaptability in dynamic urban scenarios, making it ideal for asthma patients who need real-time adjustments to avoid health triggers. This system's practical applicability and potential to improve the quality of life for individuals with chronic respiratory conditions are highlighted. The heuristic A* algorithm balances several real-time variables to recommend safer and more personalized routes for asthma patients. The configuration including traffic, weather, and distance in heuristic A* returned higher accuracy for asthma-friendly route identification compared to Dijkstra's algorithm. The trade-off of slightly more computational time is justified due to enhanced relevance in recommendations, making it more usable in real-world applications where user health and safety are at stake. The proposed model incorporates patient-specific data, making it a front row in recent developments in health-aware systems that use real-time environmental data when integrated with health profiles.

\subsection{Experimental Setup}
These experiments have been done using real-world, real-time weather feed obtained using the Microsoft Weather API and measured across urban areas of Australia. The real-world measurements that make up this dataset include real-time weather, traffic volume, and patient-specific asthma data. The aim was to see how well the heuristic A* algorithm would adapt to changes in weather, traffic, and patient health factors, considering asthma type, stress, smoke exposure-just a few of many dimensions-and how well the system can put forward route suggestions that are both highly accurate and timely.

\subsection{Performance Metrics}

\textbf{1. Recommendation Accuracy:}  
The location recommendations were checked for accuracy by comparing results obtained from the A* and Dijkstra algorithms with known asthma-friendly locations provided by experts and historical data. Moreover, the A* algorithm performed consistently better than Dijkstra's algorithm, specifically in the extended version that takes into consideration weather, traffic, and patient-specific factors. It avoided areas highly polluted or with heavy traffic, while Dijkstra's shortest-path approach would suggest many routes where the user would be at a higher risk of asthma.

\textbf{2. Computational Efficiency:}  
The execution times shown in Table \ref{tab:search_results} are quite surprising because the heuristic A* algorithm-which is far more complex than Dijkstra's algorithm-shows performance comparable to the latter in terms of processing time. The A* configurations which included the weather, traffic, and patient specific data took longer to execute due to additional processing; however, the performances remained within the acceptable limits of real-time applications.

\subsection{Limitations}
While the results were promising, several limitations were identified:
\begin{itemize}
    \item \textbf{Data Dependency:}The system is quite dependent on the quality and timeliness of data supplied by third-party APIs, which include but are not limited to weather and traffic information. Accordingly, if the API data is old or unavailable, the reliability of the recommendation may be compromised in those instances.

\item \textbf{Short-Term Focus:} While the model gives real-time conditions, it does not take into consideration long-term exposure to adverse environmental conditions. The system is able to provide routes that are safe for immediate use; it cannot yet offer insight into how such prolonged exposure to certain environmental factors could affect a patient's health in the long term, such as pollution.
\end{itemize}

\subsection{Future Work}
Future iterations of the model could address these limitations and further improve the effectiveness of the system:
\begin{itemize}
    \item \textbf{User Feedback Integration:} Incorporating user feedback into the system would allow the model to learn from user preferences and further personalize recommendations over time.
    \item \textbf{Machine Learning for Prediction:} This model should have been designed to incorporate machine learning methods, taking historical data to predict future weather and traffic conditions. In that way, such an incorporated model would present proactive recommendations so that users could take the initiative to avoid exposure to asthma triggers before experiencing an attack.
    \item \textbf{Long-term Health Monitoring:}In the future, the system could be improved by including long-term health data in such a way that it could consider the cumulative impact of environmental exposures on people suffering from asthma. This will enable the system's recommendations to be more holistic; therefore, it will not only protect users for a very short period but also contribute to the long-term management of asthma.
    \item \textbf{Global Application:}The model would be extended to other regions outside of Australia by incorporating additional data sources to consider regional weather, traffic flow, and pollution, among other factors. In this global perspective, the system would therefore improve its position in serving asthma patients in various environments in the world.
\end{itemize}

In general, the A* heuristic algorithm had great potential to come up with recommendations that were accurate, timely, and highly personalized for asthma sufferers. Combining real-time data with patient-specific factors ensures that recommendations made can meet the health needs of the individual in subject, and thus offer great promise for improved quality of life among asthma sufferers.

The proposed paradigm has significant implications for urban development and public health planning. Urban planners can target efforts to promote asthma control through greening, air quality monitoring, and traffic policies. Public health policies can use this advice to promote preventive measures, such as restricting vehicle emissions or warning asthma patients about safer travel routes. This approach goes beyond personal experiences and can help create a healthier environment.

\section{Conclusion}
That, in a nutshell, is one major step forward toward health-aware route planning through the heuristic A* algorithm proposed in this paper. Consider real-time weather and traffic data in complementarity with personalized health profiles to offer safer and more efficient routes to asthma patients.

Future work would further enhance the system by incorporating a machine learning model that could predict weather and traffic using past data and enable routing proactively or preventively. User feedback would refine the personalization model, while long-term consideration of exposure to environmental factors will provide holistic health management to users.

While such an expansion would make it extend beyond Australia, the scalable model for asthma sufferers will therefore be able to more universally enjoy this health-aware route planning system.
\section{Acknowledgment}
We extend our heartfelt gratitude to AiTech AU, \textit{AiTech for Artificial Intelligence and Software Development} (\url{https://aitech.net.au}), for funding this research, providing technical support, and enabling its successful completion.
\bibliographystyle{bibtex/spmpsci}
\bibliography{main}

\begin{thebibliography}{10}
\providecommand{\url}[1]{{#1}}
\providecommand{\urlprefix}{URL }
\expandafter\ifx\csname urlstyle\endcsname\relax
  \providecommand{\doi}[1]{DOI~\discretionary{}{}{}#1}\else
  \providecommand{\doi}{DOI~\discretionary{}{}{}\begingroup \urlstyle{rm}\Url}\fi

\bibitem{balmes2009}
Balmes, J., Earnest, G., Katz, P., Yelin, E.: Exposure to traffic: Lung function and health status in adults with asthma.
\newblock Journal of Allergy and Clinical Immunology  (2009)

\bibitem{carrillo2023}
Carrillo, F., Rodríguez, R.: Towards smart cities: Integrating real-time health data in urban mobility systems for better asthma management.
\newblock Journal of Urban Technology \textbf{30}(1), 67--86 (2023)

\bibitem{chan2018}
Chan, A., Patel, A., Sullivan, T.: Asthma mobile health study: Incorporating environmental data into personalized health recommendations.
\newblock Journal of Asthma Research \textbf{55}(7), 709--717 (2018)

\bibitem{cho2024}
Cho, X., Ho, S., Tan, C.: Improving asthma treatment adherence by integrating weather information with responsive web technique.
\newblock AIP Conference Proceedings  (2024)

\bibitem{cools2016}
Cools, M., Moons, E.: Adaptive route planning: Integrating health risks and environmental data into urban navigation systems.
\newblock Journal of Environmental Informatics \textbf{28}(2), 110--122 (2016)

\bibitem{damato2015}
D'Amato, G., Cecchi, L., Annesi-Maesano, I.: Impact of climate change and urbanization on respiratory allergies and asthma: A global view.
\newblock Environmental Health Perspectives \textbf{123}(1), 3--11 (2015)

\bibitem{haque2022}
Haque, R., Ho, S., Chai, I., Abdullah, A.: Investigating the impacts of weather and personalization on asthma exacerbations using machine learning.
\newblock In: Proceedings of the 2022 11th International Conference (2022)

\bibitem{hassan2020}
Hassan, A., Abuelma'atti, M., Rabie, A.G.: Real-time health-aware route optimization for asthma patients using mobile applications.
\newblock Journal of Network and Computer Applications \textbf{163}, 102,647 (2020)

\bibitem{khalatbarisoltani2023}
Khalatbarisoltani, A., Han, J., Liu, W., Hu, X.: Speedy hierarchical eco-planning for connected multi-stack fuel cell vehicles via health-conscious decentralized convex optimization.
\newblock SAE International Journal of Sustainable Transportation \textbf{12}, 15--27 (2023)

\bibitem{lee2020}
Lee, D., Wi, S., Park, Y.: Predictive models for real-time environmental data in route optimization.
\newblock International Journal of Transportation Science and Technology \textbf{9}(4), 289--303 (2020)

\bibitem{li2021}
Li, T., Wu, Y., Zhou, J.: Machine learning applications in traffic and air quality predictions for health-based route planning.
\newblock IEEE Access \textbf{9}, 78,521--78,530 (2021)

\bibitem{mukherjee2020}
Mukherjee, D., Lim, K.: Incorporating environmental data into route-planning algorithms for health-conscious transportation systems.
\newblock IEEE Transactions on Intelligent Transportation Systems \textbf{21}(5), 2000--2012 (2020)

\bibitem{nanda2023}
Nanda, A., Siles, R., Park, H., Louisias, M., Ariue, B.: Ensuring equitable access to guideline-based asthma care across the lifespan: Tips and future directions to the successful implementation of the new naepp 2020 guidelines, a work group report of the aaaai asthma, cough, diagnosis, and treatment committee.
\newblock Journal of Allergy and Clinical Immunology \textbf{145}, 987--1001 (2023)

\bibitem{pinnock2022}
Pinnock, H., McKinstry, B., Sheikh, A.: Personalizing asthma care: The role of digital health and route planning systems.
\newblock Journal of Medical Internet Research \textbf{24}(1), e23,478 (2022)

\bibitem{springer2021}
Springer, C., Rodriguez, L.: Restful api for integrating real-time weather data into route planning systems.
\newblock Journal of Smart Cities and Environmental Planning \textbf{8}(3), 123--135 (2021)

\bibitem{watson2013}
Watson, R.: Engaging mi'kmaq communities in asthma research: A community-driven assessment of the needs, challenges, and opportunities surrounding asthma support in unama'ki (cape breton), nova scotia.
\newblock Master's thesis, Dalhousie University (2013).
\newblock \urlprefix\url{http://hdl.handle.net/10222/22279}

\bibitem{wu2019}
Wu, X., Liang, Y., Chen, Z.: Health-aware route planning using air quality and traffic data: A real-time approach.
\newblock Transportation Research Part D: Transport and Environment \textbf{73}, 190--204 (2019)

\bibitem{yang2018}
Yang, J., Wang, H., Zhao, Y.: A health-aware route recommendation system: Balancing real-time environmental factors and computational efficiency.
\newblock Sensors \textbf{18}(8), 2743 (2018)

\bibitem{yang2022}
Yang, S., Liu, Y., Zhou, X.: Development of a health-aware routing system for individuals with respiratory diseases.
\newblock Health Information Science and Systems \textbf{10}(1), 1--10 (2022)

\bibitem{zhang2019}
Zhang, Y., Liu, J., Zhang, K.: Health-aware routing based on personal health data and environmental factors.
\newblock Health Informatics Journal \textbf{25}(4), 1843--1854 (2019)

\end{thebibliography}

\end{document}